      [1999/12/01 v1.4c Il Nuovo Cimento]
\documentclass{cimento}

\usepackage{amsmath}
\usepackage{amssymb}
\usepackage{amsthm}
\usepackage{amsfonts}

\def \de {\partial}

\def \D {\Delta}
\def \b {\beta}
\def \a {\alpha}

\def \g {\gamma}

\def \h {\eta}

\def \d {\delta}
\def \D {\Delta}

\def \r {\rho}

\def \m {\mu}
\def \n {\nu}

\def \f {\varphi}

\def \non {\nonumber}

\def \fr {\displaystyle\frac}

\def \wt {\widetilde}

\def\laq{~\raise 0.4ex\hbox{$<$}\kern -0.8em\lower 0.62
ex\hbox{$\sim$}~}
\def\gaq{~\raise 0.4ex\hbox{$>$}\kern -0.7em\lower 0.62
ex\hbox{$\sim$}~}

\def \N4 {{\CMcal N}=4}

\def \tr {\mbox{Tr}}

\def \be {\begin{equation}}
\def \ee {\end{equation}}
\def \bea {\begin{eqnarray}}
\def \eea {\end{eqnarray}}
\def \bc {\begin{cases}}
\def \ec {\end{cases}}

\newcommand{\bra}[1]{\langle{#1}|}
\newcommand{\ket}[1]{|{#1}\rangle}

\newcommand{\Tprod}[1]{T\left[{#1}\right]}

\title{Holographic scalar mesons}
\author{S.~Nicotri\from{ins:x}\thanks{stefano.nicotri@durham.ac.uk}}
\instlist{\inst{ins:x} Institute for Particle Physics Phenomenology, Department of Physics, Durham University, Durham DH1 3LE, United Kingdom}

\begin{document}

\maketitle

\begin{abstract}
A holographic description of scalar mesons is presented, in which two- and three-point function are holographically reconstructed. Mass spectrum, decay constants, eigenfunctions and the coupling of the scalar states with two pseudoscalars are found. A comparison of the results with current phenomenology is discussed.
\end{abstract}

\section{Introduction}
Many approaches and techniques have been developed to understand QCD in its
non-perturbative regime, but, up to now, no one has been completely
satisfactory, leaving strongly coupled theories still a mystery.  Recently,
the possibility to apply AdS/CFT
correspondence~\cite{Maldacena:1997re,Witten:1998qj} methods to (the large
$N$ limit of) strongly coupled gauge theories has been pointed out. This
direction, known as AdS/QCD~\cite{Polchinski:2001tt}, has been followed along
two main approaches.  The first is a top-down approach, consisting in the
attempt to obtain QCD-like theories as gravity duals of certain limits of
well defined superstring frameworks~\cite{Gursoy:2007cb}.  The second is a
phenomenological approach, consisting in building a higher dimensional model
able to describe certain relevant degrees of freedom of strong interaction,
assuming its validity as QCD dual. This is the approach followed in the
present discussion.  Many aspects of chromodynamics have been studied in this
framework, like chiral symmetry breaking~\cite{Erlich:2005qh,Karch:2006pv},
deep inelastic scattering~\cite{DIS}, deconfinement
transition~\cite{conf1,conf2} and $\bar QQ$ potential~\cite{pot1,pot2}, form
factors~\cite{pionformfactor} and
spectra~\cite{spectra,Colangelo:2008us,Colangelo:2007pt,glueball1,glueball2}.
In this paper, the model for chiral symmetry breaking introduced
in~\cite{Karch:2006pv} and, with a different aim, in~\cite{Andreev:2006vy} is
used to investigate the scalar meson sector~\cite{Colangelo:2008us}, which is
still debated nowadays, due to its features at large
$N_c$~\cite{Pelaez:2003dy}.

\section{Model}
The model is defined by the five dimensional action
\begin{equation}\label{action}
 S=-\fr{1}{k}\int d^5x\sqrt{-g}\,e^{-\Phi(z)}\tr\left\{|DX|^2+m_5^2X^2+\fr{1}{2g_5^2}\left(F_V^2+F_A^2\right)\right\}
\end{equation}
in the five dimensional Anti-de~Sitter spacetime (the bulk), defined by the
metric $g_{MN}=(R^2/z^2)\h_{MN}$, where $\h_{MN}$ is the Minkowski metric
tensor with signature $-++++$, $R$ is the AdS radius and $z$ is the fifth
{\it holographic} coordinate $0\leqslant z<\infty$. Every field is dual to a
QCD operator defined on the boundary $z=0$. $X=(X_0+S)e^{2i\pi}$ is a scalar
field, whose (negative) mass is fixed by the formula
$m_5^2R^2=(\D-p)(\D+p-4)$, where $\D$ is the dimension of the corresponding
operator and $p$ is the order of the $p$-form ({\it i.e.} $p=0$). $\pi$~is
the chiral field and $X_0=v(z)/2$ is dual to $\langle\bar qq\rangle$ and is
responsible for chiral symmetry breaking. $S=S^AT^A=S_1T^0+S_8^aT^a$ with
$T^0=(1/\sqrt{6})\mathbf{1}$ and $T^a$ the generators of $SU(3)_F$,
($A=\{0,a\}$, with $a=1,\ldots8$ ). $S^A$ is dual to the QCD operator ${\cal
O}_S^A=\bar qT^Aq$ representing the scalar
mesons. $F_V^{MN}=\de^MV^N-\de^NV^M-i\left[V^M,V^N\right]-i\left[A^M,A^N\right]$
and $F_A^{MN}=\de^MA^N-\de^NA^M-i\left[V^M,A^N\right]-i\left[A^M,V^N\right]$
are the strength tensors of the fields $V_M^a$ and $A_M^a$, obtained rotating
$A^a_{L,R}$, which are inserted to gauge, in the bulk, the global chiral
symmetry $SU(3)_L\otimes SU(3)_R$, broken to $SU(3)_V$ by $\langle\bar
qq\rangle$. $A^a_{L,R}$ are dual to the $\bar q_{L,R} \g^\m T^a q_{L,R}$
currents.  The field $\Phi(z)=c^2z^2$, which in fact defines the model, is a
non dynamical field, inserted to break the conformal symmetry in the UV
(being $c$ a mass parameter).

Assuming for QCD the validity of the AdS/CFT relation
\begin{equation}\label{adscft}
 \left\langle\exp\left[i\int d^4x\left({\cal L}+\f_0(x){\cal
 O}(x)\right)\right]\right\rangle_{QCD}=e^{iS\left[\f(x,z)\right]}
\end{equation}
where the {\it lhs} is the QCD generating functional and $\f_0(x)$ is the boundary ($z\to0$) value of the $5d$ field $\f(x,z)$, the effective action~\eqref{action} is the only ingredient needed to evaluate correlation functions.

\section{Spectrum}
To evaluate the spectrum of the scalar mesons, consider the quadratic action for the field $S^A$:
\begin{equation}
 S_{\rm eff}=-\fr{1}{2k}\int d^5x\sqrt{-g}\,e^{-\Phi}\left(g^{MN}\de_MS^A\de_NS^A+m_5^2S^AS^A\right)\,\,.
\end{equation}
%
%
%
Looking for a solution of the equation of motion which is a plane wave in the
$4d$ coordinates, $S^A(x,z)=e^{iq\cdot x}\wt S(z)$, the masses are found
solving a second order linear differential equation, whose normalizable
solutions represent the wave functions of the scalar mesons. The spectrum is discrete and the eigensystem is~\cite{scal1}
\begin{eqnarray}\label{eigen}
 m_n^2=c^2(4n+6)&~~~~~~~&\wt S_n(z)=\sqrt{\frac{2}{n+1}}c^3z^3 L_n^1(c^2z^2)
\end{eqnarray}
with $L^1_n$ the generalized Laguerre polynomials. The scalar mesons are then organized in a Regge trajectory and, fixing the parameter $c$ with the $\r$-meson mass, $c=m_\r/2$~\cite{Karch:2006pv}, they turn out to be heavier than vector mesons, with  $m_0=943$~MeV, and in good agreement with the experimental masses, considering $a_0(980)$ and $f_0(980)$ as the lightest scalar states.

\section{Two-point correlation function}
The next step is to evaluate the two-point correlation function, defined in QCD as
\begin{equation}\label{2ptqcd}
 \Pi^{AB}_{QCD}(q^2)=i\int d^4x\,e^{iq\cdot x}\bra{0}\Tprod{{\cal O}_S^A(x){\cal O}_S^B(0)}\ket{0}\,\,.
\end{equation}
Writing $\wt S(q^2,z^2)=S(q^2,z^2)\wt S_0(q^2)$, with $\wt S_0(q^2)$ the Fourier transform of the source of the operator ${\cal O_S}$ in the QCD generating functional and $S(q^2,z^2)$ a function called bulk-to-boundary propagator~\cite{Witten:1998qj}, and using~\eqref{adscft}, one can evaluate~\eqref{2ptqcd} deriving twice the effective action~\eqref{action} with respect to $\wt S_0$, obtaining:
\begin{eqnarray}\label{piads}
 \Pi^{AB}_{AdS}(q^2) & = & \d^{AB}\fr{R^3}{k}\,S(q^2,z^2)\fr{e^{-c^2z^2}}{z^3}\de_zS(q^2,z^2)\biggl|_{z=1/\n\to0}\non\\
& = & \d^{AB}\fr{4c^2R}{k}\biggl[\biggl(\fr{q^2}{4c^2}+\fr{1}{2}\biggr)\ln\left(c^2z^2\right)+\left(\g-\fr{1}{2}\right)+\fr{q^2}{4c^2}\left(2\g-\fr{1}{2}\right)\\
&& +\biggl(\fr{q^2}{4c^2}+\fr{1}{2}\biggr)\psi\left(q^2/4c^2+3/2\right)\biggr]\biggl|_{z=1/\n}\,\,.\non
\end{eqnarray}
This function has poles at $-q^2_n=m_n^2=c^2(4n+6)$, in agreement with~\eqref{eigen}, with residues $F_n^2=16Rc^4(n+1)/k$ corresponding to the decay constants of the scalar mesons. The factor $R/k$ can be fixed with a comparison of~\eqref{piads} with the known QCD result, obtaining $R/k=N_c/(16\pi^2)$. The AdS prediction $F_0=0.08$~GeV$^2$ can be compared to QCD determinations $F_{a_0}=(0.21\pm0.05)$~GeV$^2$ and $F_{f_0}=0.18\pm0.015$~GeV$^2$~\cite{Gokalp,DeFazio}, showing a difference of about a factor of two.

\section{Three-point correlation function and interaction with two pseudoscalars}

The three-point correlation function describing the interaction between a scalar meson and two pseudoscalars is defined in QCD by
%
%
%
\begin{equation}\label{3ptqcd}
 \Pi^{abc}_{QCD\,\a\b}(p_1,p_2)=d^{abc}\fr{p_{1\a}\, p_{2\b}}{p_1^2\,
 p_2^2}f_\pi^2\sum_{n=0}^\infty\fr{F_n\,g_{S_nPP}}{q^2+m_n^2} 
\end{equation}
with $q=-(p_1+p_2)$, $f_\pi$ the pion decay constant and $g_{S_nPP}$ the coupling.

To evaluate it on the AdS side, consider the corresponding interaction term in~\eqref{action}:
\begin{equation}
 S_{\rm eff}^{SPP}=-\fr{R^3}{k}\int
 d^5x\, \fr{e^{-\Phi(z)}}{z^3}\,v(z)\left[\fr{2}{\sqrt{6}}\,S_1(\de\psi)^2+d^{abc}S_8^a\h^{MN}\left(\de_M\psi^b\right)\left(\de_N\psi^c\right)\right]
\end{equation}
where $A_M=A_{\perp M}+\de_M\phi$ and $\psi^a=\phi^a-\pi^a$ is the pseudoscalar dual field. Differentiating with respect to the sources, the coupling for the $n=0$ state is given by
\begin{equation}\label{gspp}
 g_{S_0PP}=\fr{m^2_{S_0}Rc\sqrt{N_c}}{4\pi f_\pi^2}\int_0^\infty du\,e^{-u^2}v(u)
\end{equation}
with $u=cz$. The numerical result is of ${\cal O}(10)$~MeV, at odds with
experimental values $g_{a_0\h\pi}=12\pm6~\mbox{GeV}$ and phenomenological
determinations
$g_{f_0K^{+}K^{-}}\simeq7~\mbox{GeV}$~\cite{Colangelo:2003jz}. This is an
issue of the model and it is due to the fact that the integral
in~\eqref{gspp} is dominated by the small quark mass parameter. This is
related to the difficulty of the model in correctly reproducing both
spontaneous and explicit chiral symmetry breaking, difficulty already
analyzed in~\cite{Karch:2006pv}.


\acknowledgments
The author  gratefully acknowledges an Early Stage Researcher
position supported by the EU-RTN Programme, Contract No.
MRTN--CT-2006-035482, \lq\lq Flavianet\rq\rq.

It is a pleasure to thank P.~Colangelo, F.~De~Fazio, F.~Giannuzzi and F.~Jugeau for collaboration.

\end{document}